\documentclass[10pt,conference]{IEEEtran}
\usepackage{amssymb,amsthm,amsmath,url,color}

\newcommand{\FF}{\mathbb{F}}

\newtheorem{lem}{Lemma}
\newtheorem{cor}{Corollary}
\newtheorem{prop}{Proposition}
\newtheorem{defn}{Definition}
\newtheorem{exmp}{Example}

\title{Rank Weight Hierarchy  \\ of Some Classes of Cyclic Codes}
\author{%
\IEEEauthorblockN{J\'er\^ome Ducoat and Fr\'ed\'erique Oggier}
\IEEEauthorblockA{Division of Mathematical Sciences\\
              Nanyang Technological University\\
              Singapore\\
              Email:jducoat@ntu.edu.sg,frederique@ntu.edu.sg}
}

\begin{document}
\maketitle

\begin{abstract}
We study the rank weight hierarchy, thus in particular the rank metric, of cyclic codes over the finite field $\FF_{q^m}$, $q$ a prime power, $m \geq 2$. We establish the rank weight hierarchy for $[n,n-1]$ cyclic codes and characterize $[n,k]$ cyclic codes of rank metric 1 when (1) $k=1$, (2) $n$ and $q$ are coprime, and (3) the characteristic $char(\FF_q)$ divides $n$. Finally, for $n$ and $q$ coprime, cyclic codes of minimal $r$-rank are characterized, and a refinement of the Singleton bound for the rank weight is derived.
\end{abstract}

%
%
%

\section{Introduction}

Let $\FF_q$ be the finite field with $q$ elements, $q$ a prime power, and consider its field extension $\FF_{q^m}$, $m\geq 1$. 
Let $C$ be an $[n,k]$ linear code over $\FF_{q^m}$. For $c=(c_1,\ldots,c_n)$ a codeword of $C$, we denote by $\lambda(c)$ the matrix obtained by writing every $c_i$ as a vector in a $\FF_q$-basis of $\FF_{q^m}$:
\[
\lambda(c)=
\begin{bmatrix}
c_{1,1} & \ldots & c_{n,1} \\
\vdots   &       & \vdots \\
c_{1,m} & \ldots & c_{n,m}
\end{bmatrix}.
\]
The rank weight of the codeword $c$ is defined \cite{Gabidulin} as its $\FF_q$-rank, that is as the rank of $\lambda(c)$, and the rank distance $d_R(C)$ of the code $C$ is 
\[
d_R(C)=d_1(\lambda(C))=\underset{c\in C}{\underset{c \neq 0}{\min}}~\textsf{rk}(\lambda(c)).
\]
The notion of rank distance (or rank metric) of a code has been extended to that of a rank weight hierarchy $d_1(\lambda(C)),\ldots,d_k(\lambda(C))$ in \cite{OS12,KMU13}. More precisely, it was shown in \cite{D13} that a refinement of the definition of \cite{OS12} gives a definition equivalent to that of \cite{KMU13}, namely:
\begin{defn}\label{def:dr}
Let $1\leq r \leq k$ and $C$ be an $[n,k]$ linear code over $\FF_{q^m}$. The $r^{\textrm{th}}$ rank weight of $C$ is 
\[d_r(\lambda(C))=\underset{\dim(C\cap V) \geq r}{\underset{V\in\Gamma(\FF^n_{q^m})}{\min}}\dim V,
\]
where $\Gamma(\FF^n_{q^m})=\{V \subset \FF^n_{q^m}~|~V^q=V\}$, with $V^q=\{(c_1^q,\ldots,c_n^q)~|~c\in V\}$.
\end{defn}
Set $D^*=\sum_{j=0}^{m-1}D^{q^j}$. When $n\leq m$, we also have
\[d_r(\lambda(C))=\underset{\dim D=r}{ \underset{D \subset C}{\min}} \underset{c\in D^*}{\max} ~\textsf{rk}(\lambda(c)),\]

The motivation (for both \cite{OS12,KMU13}) to introduce this rank weight hierarchy is to study the equivocation of wiretap codes for network coding.


Basic properties of the rank weight hierarchy are known: 
\begin{itemize}
\item
The monotonicity property holds \cite{KMU13}: 
\begin{equation}\label{eq:mon}
d_1(\lambda(C))<\ldots<d_k(\lambda(C)) \leq n.
\end{equation} 
\item
There is a generalized Singleton bound \cite{KMU13}: 
\begin{equation}\label{eq:sing}
d_r(\lambda(C))\leq n-k+r,
\end{equation} 
and in the case of the rank weights, the Griesmer bound is the same as the generalized Singleton bound \cite{D13}.
\end{itemize} 
\begin{defn}
An $[n,k]$ linear code $C$ is {\em $r$-MRD} (maximum rank distance) if $d_r(\lambda(C))=n-k+r$, reaching (\ref{eq:sing}).
\end{defn}
Finally, the following is also known \cite{D13}: 
\begin{prop}
\label{p1}
Let $C^\perp$ be the dual code of $C$. Then 
\[
\begin{array}{c} 
\{d_r(\lambda(C)) | 1\leq r\leq k\}\\  
\sqcup \{n+1-d_s(\lambda(C^\perp)) |1\leq s\leq n-k\}=\{1,...,n\}.
\end{array}
\]
\end{prop}

In this paper, we are interested in the rank weight hierarchy of cyclic codes.
Let $C$ be a cyclic code of length $n$ over $\FF_{q^m}$ with generator polynomial $g(x)$ of degree $s$. Then $C$ is an ideal of $\FF_{q^m}[x]/(x^n-1)$, and $g(x)$ divides $x^n-1$. The dimension of $C$ is $k=n-s$, $1 \leq s \leq n-1$.

The rank distance of cyclic codes of dimension $k=1,2$ has been studied in \cite{SR03}, where instead of computing the rank of cyclic codes directly, the authors computed the discrete Fourier transform of the cyclic codewords, and obtained characterization of the rank distance in the Fourier domain.

Our results on the rank weight hierarchy of cyclic codes are as follows.
The rank weight hierarchy of $[n,n-1]$ cyclic codes is established in Section \ref{sec:first}. 
The rank distance of $[n,1]$ cyclic code is computed in Section \ref{sec:d1}. In particular, we recover the rank metric of $[n,1]$ codes discussed in \cite{SR03}.
Cyclic codes of dimension $k$ else than $1$ and $n-1$ are discussed in Section \ref{sec:kcoprimeq} and \ref{sec:kchardivides}, where codes of rank weight 1 are characterized respectively for the case when the length $n$ is coprime to $q$ and the characteristic $char(\FF_q)$ divides $n$.
Finally, in Section \ref{sec:higher}, cyclic codes of minimal $r$-rank are characterized, and a refinement of the Singleton bound for the rank weight is derived,
under the assumption that $n$ and $q$ are coprime.

%
%
%
\section{Rank Weight Hierarchy of $[n,n-1]$ Cyclic Codes}
\label{sec:first}

A cyclic code of dimension $n-1$ has a generator polynomial of degree $1$.
Its generator matrix $G$ is by definition
\begin{equation}\label{eq:gen}
\begin{bmatrix}
g_0 & 1 & 0   & \cdots & 0 \\
0   & g_0 & 1 &\ddots & \vdots \\
\vdots    &  \ddots  & \ddots  &\ddots & 0 \\
0&\cdots & 0 & g_0 & 1      
\end{bmatrix}.
\end{equation}

\begin{lem}
\label{l1}
If the generator polynomial $g(x)$ of a cyclic code $C$ over $\FF_{q^m}$ has degree 1, then the minimum rank distance $d_1(\lambda(C))$ is 1.
\end{lem}

\begin{IEEEproof}
Write $g(x)=x+g_0 \in \FF_{q^m}[x]$.
Using the generator matrix $G$ defined in (\ref{eq:gen}) of $C$,
a direct computation shows that a codeword of $C$ is of the form
\[
[c_0,c_1,c_2,\ldots,c_{n-2}]G=
[g_0c_0,c_0+g_0c_1,c_1+g_0c_2,\ldots ,c_{n-2}].
\]

To show that a code has rank distance $1$, we only need to exhibit a codeword with rank weight $1$. 

Let $l$ be the degree of the minimal polynomial of $g_0$ over $\FF_q$. 
Since $-g_0$ is a root of the polynomial $x^n-1$ over $\FF_q$ and since $1$ is a trivial root, we have $l\leq n-1$. Let $\lambda_0,...,\lambda_{l-1}\in \FF_q$ be the coefficients of this minimal polynomial, i.e., elements of $\FF_q$ such that 
$$
\lambda_l g_0^l+\lambda_{l-1}g_0^{l-1}+ \cdots+ \lambda_0=0,~\lambda_l=1.
$$
Take for $0\leq i\leq n-2-l$,  $c_i=0$ and for $n-1-l\leq i\leq n-2$
$$c_{i}= (-1)^{n-i} \underset{j=0}{\overset{n-2-i}{\sum}}\lambda_{l-n+2+i+j}g_0^{j}. $$
Since the $i$th coefficient of a codeword is $c_{i-2}+g_0c_{i-1}$, we obtain a codeword whose $i$th coefficient is zero, for $1\leq i\leq n-1-l$. 
For $n-l\leq i\leq n$, the $i$th coefficient is
\[
\begin{array}{ll} 
 & c_{i-2}+g_0c_{i-1} \\ 
=& (-1)^{n-(i-2)} \underset{j=0}{\overset{n-2-(i-2)}{\sum}}\lambda_{l-n+2+(i-2)+j}g_0^{j}\\ 
 & +g_0 (-1)^{n-(i-1)} \underset{j=0}{\overset{n-2-(i-1)}{\sum}}\lambda_{l-n+2+(i-1)+j}g_0^{j} \\ 
=& (-1)^{n-i} \underset{j=0}{\overset{n-i}{\sum}}\lambda_{l-n+i+j}g_0^{j} \\
 & + (-1)^{n-i+1} \underset{j=0}{\overset{n-1-i}{\sum}}\lambda_{l-n+1+i+j}g_0^{j+1} \\ 
=& (-1)^{n-i}\big( \lambda_{l-n+i}+ \underset{j=1}{\overset{n-i}{\sum}}\lambda_{l-n+i+j}g_0^{j} - \underset{j=1}{\overset{n-i}{\sum}}\lambda_{l-n+i+j}g_0^{j} \big)\\ 
=& (-1)^{n-i} \lambda_{l-n+i},  
\end{array}
\]
and the $n$th coefficient is $c_{n-2}=1$, showing that $c_i\in\FF_q$, $0\leq i \leq n-2$.
\end{IEEEproof}

Using Proposition \ref{p1}, it is enough to determine the rank hierarchy of the dual code $C^\perp$ to know completely the rank hierarchy of $C$.

Recall from (\ref{eq:gen}) the $(n-1)\times n$ generator matrix $G$ of $C$. 
The parity check matrix of the dual code $C^\perp$ of $C$ is then $G^t$ : a vector $d=[d_1,\ldots,d_n]\in \FF_{q^m}^n$ is in $C^\perp$ if and only if 
$dG^t=0$, which is equivalent to :
$$\begin{cases}
g_0d_1+d_2=0\\
g_0d_2+d_3=0\\
\hspace{1.7cm}\vdots\\
g_0d_{n-1}+d_n=0.
\end{cases}$$

Hence, $C^\perp$ is the $1$-dimensional vector space on $\FF_{q^m}$ generated by the vector $$[1,-g_0,\ldots, (-g_0)^{n-1}].$$

Therefore, the rank weight of this vector is the dimension of the $\FF_q$-vector space generated by the family $\{(-g_0)^i\}_{0\leq i\leq n-1}$: it is equal to the degree of the minimal polynomial of $-g_0$ (equivalently of $g_0$) over $\FF_q$. 

Hence, we have the following result :
\begin{cor}
Keeping the notation as above, we have :
\begin{enumerate}
\item for $1\leq r\leq n-[\FF_q(g_0):\FF_q]$, $$d_r(\lambda(C))= r.$$
\item for $n+1-[\FF_q(g_0):\FF_q]\leq r\leq n-1$, $C$ is a $r$-MRD code.
\end{enumerate}
\end{cor}

\begin{IEEEproof}
This follows from the monotonicity property (\ref{eq:mon}), from Proposition \ref{p1} and from the above computation of the first rank distance of $C^\perp$.
\end{IEEEproof}

%
%
%
\section{Rank Weight of $[n,1]$ Cyclic Codes}
\label{sec:d1}

Let $g(x)=g_0+g_1x+\ldots+g_{n-2}x^{n-2}+x^{n-1}$ be the generator polynomial of an $[n,1]$ cyclic code $C$, and let $h(x)=x+h_0$ be its check polynomial, satisfying
\[
g(x)h(x)=x^n-1.
\]
Then $g_0h_0=-1$ and $h(x)=x-g_0^{-1}$. 
The dual code $C^\perp$ of $C$ has dimension $n-1$, and generator polynomial
\[
h_0^{-1}xh(x^{-1})=-g_0x(x^{-1}-g_0^{-1})=x-g_0.
\]
The computations of Section \ref{sec:first} tell that the dual of $C^\perp$, that is 
$C$, is the 1-dimensional vector space on $\FF_q$ generated by 
\[
[1,g_0,g_0^2,\ldots,g_0^{n-1}]
\]
and thus: 
\begin{lem}\label{lem:d1}
Let $C$ be an $[n,1]$ cyclic code with generator polynomial $g(x)$.
Then its rank weight is $[\FF_q(g(0)):\FF_q]$. 
\end{lem}

As a consequence, we obtain the rank distance of the four cyclic codes of dimension 1 computed in \cite{SR03-2}. 
\begin{exmp}\rm
Consider a primitive length cyclic code $C$ over $\FF_{2^4}$, that is of length 
$n=|\FF_{2^4}|-1=15$, and dimension $k=1$. Then 
\[
x^{15}-1=\prod_{i=0}^{14}(x-\alpha)
\]
where $\alpha$ is a primitive element of $\FF_{2^4}^*$. Let $g(x)$ be the generator polynomial of $C$, whose constant coefficient $g_0$ may be any element of $\FF_{2^4}^*$. 
Since $\alpha$ is of order 15, $\alpha^{3i}$ is of order 5, $i=1,2,3,4$, $\alpha^{5i}$ is of order $3$, $i=1,2$. Thus when $g_0=\alpha^5$ or $\alpha^{10}$, $C$ has rank distance 2 (the minimum polynomial of $g_0$ is $x^2+x+1$), which corresponds to Example 3 of \cite{SR03-2}. Otherwise, the minimum polynomial of $g_0$ has degree 4, and the code has rank distance 4, as was computed in Example 2 of \cite{SR03-2}. The other examples of \cite{SR03-2} are computed similarly.   
\end{exmp}

%
%
%

\section{Characterization of $[n,k]$ Cyclic Codes of Rank Weight $1$ when the length is coprime to $q$}
\label{sec:kcoprimeq}
\subsection{Case I: the Generator Polynomial is Split.}

Let $C$ be an $[n,k]$ cyclic code. In this section, we assume that $n$ and $q$ are coprime, which implies that all the roots of $x^n-1$ are simple. We denote by $\alpha_1,...,\alpha_\nu$ those roots belonging to $\FF_{q^m}$ (they are pairwise distinct). Let $g(x)$ be the generator polynomial of $C$. We assume that $g(x)$ is split in $\FF_{q^m}[x]$. Since the dimension of $C$ is $k$, $g(x)$ has degree $n-k$. Since $g(x)$ divides $x^n-1$, we have $n-k\leq \nu$ and up to re-ordering the $\alpha_i$, we may assume that 
\[
g(x)=\underset{1\leq j\leq n-k}{\prod}(x-\alpha_j).
\]

Let $G$ be the generator matrix of $C$. Then a codeword 
\[
[c_0,c_1,\ldots,c_{k-1}]G,
\]
is written in terms of polynomial as 
\[
c(x)g(x),~c(x)=c_0+c_1x+\ldots+c_{k-1}x^{k-1}.
\]

Since $g(x)$ is of degree $\leq n-k$, we indeed get a polynomial of degree $\leq n-1$, whose $n$ coefficients correspond to one codeword.
Thus any codeword can be written as
\[
c(x)\prod_{1\leq j\leq n-k}(x-\alpha_j).
\]
Moreover, a code $C$ has rank weight $1$ if and only if there exists a codeword with coefficients in $\FF_q$, which means here that the corresponding polynomial $c(x)\prod_{1\leq j\leq n-k}(x-\alpha_j)$ lives in $\FF_q[x]$.

Recall that $g(x)$ is split with simple roots $\alpha_1,\ldots,\alpha_{n-k}$. Up to re-ordering the roots, let $m_1\geq 1$ be such that $\alpha_1,\ldots,\alpha_{m_1}$ are roots of the minimal polynomial $\mu_{\alpha_1}(x)$ of $\alpha_1$ over $\FF_q$, let $m_2\geq m_1+1$ be such that $\alpha_{m_1+1},\ldots,\alpha_{m_2}$ are roots of the minimal polynomial $\mu_{\alpha_{m_1+1}}(x)$ of $\alpha_{m_1+1}$ over $\FF_q$,... and let $m_s\geq m_{s-1}+1$ be such that $\alpha_{m_{s-1}+1},\ldots,\alpha_{m_s}=\alpha_{n-k}$ are roots of the minimal polynomial $\mu_{\alpha_{m_{s-1}+1}}(x)$ of $\alpha_{m_{s-1}+1}$ over $\FF_q$. 

Now, for any $1\leq r\leq s-1$, $\mu_{\alpha_{m_r+1}}(x)$ divides $x^n-1$. Since $\FF_{q^m}/\FF_q$ is a Galois extension, $\mu_{\alpha_{m_r+1}}(x)$ has a root in $\FF_{q^m}$ and is irreducible over $\FF_q$, then $\mu_{\alpha_{m_r+1}}(x)$ splits over $\FF_{q^m}$ : there exists a subset (maybe empty) $J_r\subset \{\alpha_{n-k+1},...,\alpha_\nu\}$\[\mu_{\alpha_{m_r+1}}(x)=\underset{m_r+1\leq t\leq m_{r+1} }{\prod}(x-\alpha_t) \underset{j\in J_r}{\prod} (x-\alpha_j).\]
Note that any two $J_r$ are disjoint.

From now on, we will use the following terminology :
\begin{defn}
We denote by $\eta_q(C)$ the quantity $$\underset{1\leq r\leq s-1}{\sum} [\FF_q(\alpha_{m_r+1}):\FF_q].$$ 
\end{defn}

Note that $\eta_q(C)$ only depends on the factorization of $g(x)$ in $\FF_{q^m}[x]$ and is then completely determined by $C$. In general, we have $\eta_q(C)\leq n$.

\begin{prop}
\label{p2}
Let $C$ be an $[n,k]$ cyclic code over $\FF_{q^m}$, with $n$ coprime with $q$. Assume that the generator polynomial $g(x)$ is split in $\FF_{q^m}[x]$. Keeping the notation introduced above, $C$ has rank weight $1$ if and only if $\eta_q(C)\leq n-1$.
\end{prop}

\begin{IEEEproof}
Assume first that $\eta_q(C)\leq n-1$. Using the previous description of codewords of $C$, we set 
\[c(x)=\underset{1\leq r\leq s-1}{\prod}\underset{j\in J_r}{\prod}(x-\alpha_{j}) .\]
Then the polynomial $c(x)g(x)$ has coefficients in $\FF_q$ since
\[\begin{aligned} 
& \underset{1\leq r\leq s-1}{\prod}\underset{j\in J_r}{\prod} (x-\alpha_j) \underset{1\leq j\leq n-k}{\prod}(x-\alpha_j) \\ &= \underset{1\leq r\leq s-1}{\prod}\underset{j\in J_r}{\prod} (x-\alpha_j)\underset{1\leq r\leq s-1}{\prod}\underset{m_r+1\leq t\leq m_{r+1} }{\prod}(x-\alpha_t) \\ &= \underset{1\leq r\leq s-1}{\prod} \mu_{\alpha_{m_r+1}}(x).\end{aligned}
\]
Since this polynomial has degree $\eta_q(C)\leq n-1$, $c(x)g(x)$ corresponds to a codeword of $C$ with coefficients in $\FF_q$. 

We now show the converse. Assume that $C$ has rank weight $1$, i.e. that there exists a polynomial $c(x)$ with degree $\leq k-1$ such that $c(x)g(x)$ has coefficients in $\FF_q$. Since, for $1\leq r\leq s-1$, $\alpha_{m_r+1}$ is a root of $c(x)g(x)\in \FF_q[x]$, its minimal polynomial $\mu_{\alpha_{m_r+1}}(x)$ divides $c(x)g(x)$ in $\FF_q[x]$. This being true for every $1\leq r\leq s-1$ and the polynomials $\mu_{\alpha_{m_r+1}}$ being pairwise coprime, the polynomial \[\underset{1\leq r\leq s-1}{\prod} \mu_{\alpha_{m_r+1}}(x)\] divides $c(x)g(x)$ in $\FF_q[x]$. Taking the degrees, we get the desired inequality : $\eta_q(C)\leq n-1$.  
\end{IEEEproof}

\begin{cor}
\label{c2}
Let $C$ be an $[n,k]$ cyclic code with length $n$ dividing $q^m-1$. Then
$C$ has rank weight $1$ if and only if $\eta_q(C)\leq n-1$. 
\end{cor}

\begin{IEEEproof}
Indeed, since $n|q^m-1$, the polynomial $x^n-1$ is split in $\FF_{q^m}[x]$ and we apply Proposition \ref{p2}.
\end{IEEEproof}

Recall that when $n=q^m-1$, a cyclic code is called {\em primitive}, or of {\em primitive length}.

\begin{cor}
\label{c3}
Let $C$ be an $[n,n-k]$ cyclic code with primitive length. Then $C$ has rank weight $1$ if $km\leq q^m-2$. 
\end{cor}

\begin{IEEEproof}
We have $\eta_q(C)=\underset{1\leq r\leq s-1}{\sum} [\FF_q(\alpha_{m_r+1}):\FF_q]$, $s-1\leq \deg g(x)=k$ and $[\FF_q(\alpha_{m_r+1}):\FF_q]\leq m$ for all $1\leq r\leq s-1$, since $\alpha_{m_r+1}\in\FF_{q^m}$. Corollary \ref{c3} then follows from Corollary \ref{c2}.
\end{IEEEproof}


Applying Corollary \ref{c3} when $k=2$, since $m\geq 2$, the only case for which $2m > q^m-2$ is when $q=2$ and $m=2$, that is we have a $[3,1]$ cyclic code over $\FF_4$. Using Lemma \ref{lem:d1}, its rank weight is $[\FF_2(g(0)):\FF_2]$, where $g(x)$ is a polynomial of degree of $2$ which divides $x^3-1=(x-1)(x^2+x+1)=(x-1)(x-\alpha)(x-\alpha+1)$. The rank weight is thus $1$ if $g(x)=x^2+x+1$ and $2$ otherwise. Note that we find the same result using Corollary \ref{c2} and the Singleton bound.

%
%
%

\subsection{Case II: the Generator Polynomial is not Split.}
\label{sec:k2}
Let $C$ be an $[n,k]$ cyclic code such that $n$ and $q$ are coprime. Let $g(x)\in \FF_{q^m}[x]$ be the generator polynomial of $C$. Let now $m'$ be a multiple of $m$ such that $\FF_{q^{m'}}$ is a splitting field of $g(x)$. Since $n$ and $q$ are coprime, as before, the roots of $x^n-1$ (and then of $g(x)$) are all simple (in $\FF_{q^{m'}}$). 

We extend the definition of $\eta_q(C)$ as follows :
\begin{defn}
\[\eta_q(C)=\eta_q(C\otimes_{\FF_{q^m}}\FF_{q^{m'}}).\]
\end{defn}

As before, let $\alpha_1,...,\alpha_\nu$ be roots of $g(x)$ in $\FF_{q^{m'}}$ such that every root of $g(x)$ in  $\FF_{q^{m'}}$ is conjugate to exactly one $\alpha_i$, for $1\leq i\leq \nu$. Let $g(x)=g_1(x)\cdots g_{\nu'}(x)$ be the factorization in $\FF_{q^m}[x]$ into irreducible polynomials. 

\begin{lem} 
\label{l3}
We have $\nu'=\nu$ and up to re-ordering the roots $\alpha_i$, for every $1\leq i\leq \nu$, $g_i(x)$ is the minimal polynomial of $\alpha_i$ over $\FF_{q^m}$.
\end{lem}

\begin{IEEEproof} 
The polynomial $g_i(x)$ also splits in $\FF_{q^{m'}}[x]$ so has a root $\alpha$ : therefore, the minimal polynomial of $\alpha$ over $\FF_{q^m}$ divides $g_i(x)$  in $\FF_{q^m}[x]$ and since $g_i(x)$ is irreducible and since both are monic, $g_i(x)$ is the minimal polynomial of $\alpha$. Yet, $\alpha$ is conjugate to one of the $\alpha_j$, say $j=i$ if we re-order correctly.
\end{IEEEproof}

From Lemma \ref{l3}, we can deduce that, for all $1\leq i\leq \nu$,  the minimal polynomial $\mu_{\alpha_i}(x)$ of $\alpha_i$ over $\FF_q$ can be factorized in $\FF_{q^m}[x]$ as : \[\mu_{\alpha_i}(x)=\underset{j\in J_i}{\prod} g_j(x)h_i(x),\]
where 
\[
J_i=\{j\in \{1,...,\nu\}| g_j(x) \textrm{ and } \mu_{\alpha_i}(x) \textrm{ have a common root} \}
\] 
and $h_i(x)$ is a factor of $h(x)=\frac{x^n-1}{g(x)}$ in $\FF_{q^m}[x]$. Note that some of the $\alpha_i$ may have the same minimal polynomial $\mu_{\alpha_i}(x)$ over $\FF_q$, so that two sets $J_i$ and $J_{i' }$ are equal or have empty intersection; simultaneously, the two corresponding polynomials $h_i(x)$ and $h_{i'}(x)$ are either equal, either pairwise coprime.

\begin{prop}
\label{p3}
Let $C$ be an $[n,k]$ cyclic code over $\FF_{q^m}$, $n$ and $q$ being coprime. Then $C$ has rank weight $1$ if and only if  $\eta_q(C) \leq n-1$.\end{prop}

\begin{IEEEproof}
Assume that $\eta_q(C)\leq n-1$. Keeping notation above, let $I\subset\{1,...,\nu\}$ be a set of indices such that the subsets $J_i$ are pairwise distinct and that for any $1\leq j\leq \nu$, $h_j(x)=h_i(x)$ for some $i\in I$. We then set $c(x)=\underset{i\in I}{\prod}h_i(x)\in \FF_{q^m}[x]$. Then we have 
\[
\begin{aligned} c(x)g(x) &=\underset{i\in I}{\prod} h_i(x) \cdot  g_1(x) \cdots g_{\nu}(x) \\ &= \underset{i\in I}{\prod} h_i(x) \cdot \underset{i\in I}{\prod} \underset{j\in J_i}{\prod} g_j(x) =  \underset{i\in I}{\prod} \mu_{\alpha_i}(x).
\end{aligned}
\]
The latter product has factors lying in $\FF_q[x]$ and has degree $\eta_q(C)\leq n-1$. Therefore, the corresponding codeword of $C$ has coefficients in $\FF_q$ and $C$ has rank weight $1$.

Conversely, assume that $C$ has rank weight $1$. Then there exists a polynomial $c(x)\in \FF_{q^m}[x]$ with degree $\leq k-1$ such that $c(x)g(x)\in \FF_q[x]$. But then $c(x)\in \FF_{q^{m'}}[x]$, we can consider the corresponding codeword as an element of $C\otimes_{\FF_{q^m}}\FF_{q^{m'}}$. Using now Proposition \ref{p2}, we get that $\eta_q(C)\leq n-1$.
\end{IEEEproof}

%
%
%
\section{$[n,k]$ Cyclic Codes of Rank Weight $1$ when $char(\FF_q)$ divides $n$}
\label{sec:kchardivides}

Set $p=\textrm{char}(\FF_q)$ and let $n=\widetilde n \cdot p^v$. Then 
$x^n-1=(x^{p^v})^{\widetilde n}-1$, so it has a factorization in $\FF_{q^m}[x]$ of the following form :
\[x^n-1= \underset{i=1}{\overset{\nu}{\prod}}\big(g_i(x)\big)^{p^v}=\underset{i=1}{\overset{\nu}{\prod}}g_i(x^{p^v})\] where for all $1\leq i\leq \nu$, $g_i(x)\in \FF_{q^m}[x]$ is irreducible. Let $C$ be an $[n,k]$ cyclic code over $\FF_{q^m}$ and let $g(x)$ be its generator polynomial. There exists a subset $J\subset\{1,...,\nu\}$ such that \[g(x)=\underset{i\in J}{\prod} \big(g_i(x)\big)^{l_i},\] where $l_i\leq p^{v}$ for all $i\in J$. In this section, we assume that for all $i\in J$, $l_i=p^{v_i}$ for some $v_i\leq v$. Set now \[\widetilde g(x)=\underset{i\in J}{\prod} g_i(x).\]
Then $\widetilde g(x)$ divides $x^{\widetilde n}-1$ in $\FF_{q^m}[x]$. Let $\widetilde k=\widetilde n-\underset{i\in J}{\sum} \deg (g_i(x))$ and let $\widetilde C$ be the $[\widetilde n, \widetilde k]$ cyclic code over $\FF_{q^m}$ with generator polynomial $\widetilde g(x)$.

\begin{prop}
\label{p4}
Let $C$ be an $[n,k]$ cyclic code over $\FF_{q^m}$ with generator polynomial $g(x)$ of the form ${\prod}_{i\in J} \big(g_i(x)\big)^{p^{v_i}}.$ Keeping the notation above, if $\eta_q(\widetilde C)\leq \widetilde n-1$, then $C$ has rank weight $1$.
\end{prop}

\begin{IEEEproof}
Assume that $\eta_q(\widetilde C)\leq \widetilde n-1.$ From Proposition \ref{p3}, $\widetilde C$ has rank weight $1$, so there exists some polynomial $\widetilde c(x)\in \FF_{q^m}[x]$ such that $\widetilde c(x)\widetilde g(x)$ has coefficients in $\FF_q[x]$ and degree $\leq \widetilde n-1$. If $v_0=\max_{i\in J}(v_i)$, we set $c(x)=\widetilde c(x^{p^{v_0}})$. Then $c(x)g(x)$ is a polynomial with coefficients in $\FF_q[x]$ and degree   \[\begin{aligned} \deg c(x)g(x) &= p^{v_0}(\widetilde k-1)+\deg g(x)\\ &\leq p^{v_0}(\widetilde k-1+\deg \widetilde g(x))\\ &\leq p^v (\widetilde n-1)\leq n-1.\end{aligned}\] Hence, the corresponding codeword has rank weight $1$.
\end{IEEEproof}

%
%
%

\section{Higher rank weights and dual codes}
\label{sec:higher}

\subsection{Characterization of Cyclic Codes with Minimal $r$-Rank}
As a natural generalization of Proposition \ref{p3}, we have :
\begin{prop}
\label{p5}
Let $C$ be an $[n,k]$ cyclic code over $\FF_{q^m}$ with $n$ coprime with $q$ and let $1\leq r\leq k$. Then $d_r(\lambda(C))=r$ if and only if $\eta_q(C) \leq n-r$. 
\end{prop}

\begin{IEEEproof}
Assume first that $\eta_q(C)\leq n-r$. Then, taking the polynomial $c(x)$ defined in the proof of Proposition \ref{p3}: \[c(x)=\underset{i\in I}{\prod}h_i(x),\] we set, for every $0\leq u\leq r-1$, $c_u(x)=x^uc(x)$. Then, for all $0\leq u\leq r-1$, $c_u(x)g(x)$ is a polynomial lying in $\FF_q[x]$ with degree $\leq n-1$. It then corresponds to a codeword $c_u$ with rank weight $1$. Moreover, the subspace $V$ of $C$ generated by the $c_u$'s has dimension exactly $r$ (for all $0\leq u\leq r$, the polynomial $c_u(x)g(x)$ has degree $n-r+u$, so the family of the codewords $c_u$ is linearly independent) and $V$ belongs to $\Gamma(\FF_{q^m}^n)$, as in Definition \ref{def:dr} (since the basis vectors $c_u$ lie in $\FF_q^n$). Therefore, $d_r(\lambda(C))\leq \dim V=r$. Moreover, as a direct consequence of the monotonicity property \cite{KMU13}, $r\leq d_r(\lambda(C)$ and we get the desired equality.

Conversely, assume that $d_r(\lambda(C))=r$. Then there exists a subspace $V\in \Gamma(\FF_{q^m}^n)$ with dimension $r$ such that $\dim(V\cap C)\geq r$. Hence, $V\subset C$. Moreover, we know from \cite{S90}, that $V$ has a basis of vectors having coefficients in $\FF_q$ : there exists some polynomials $c_1(x),...,c_r(x)\in \FF_{q^m}[x]$ with degree $\leq k-1$ such that $c_i(x)g(x)\in \FF_q[x]$ and the family $\{c_i(x)g(x)|1\leq i\leq r\}$ is linearly independent over $\FF_{q^m}$. Therefore, there exists a non-zero polynomial $c(x)\in \FF_{q^m}[x]$ with degree $\leq k-r$ lying in the subspace spanned by the $c_i(x)g(x)$ over $\FF_q$. Keeping the notation introduced in the proof of Proposition \ref{p3}, the minimal polynomial of any root $\alpha$ (say in an algebraic closure of $\FF_q$) over $\FF_q$ divides $c(x)g(x)$, hence \[\big( \underset{i\in I}{\prod} \mu_{\alpha_i}(x) \big) | c(x)g(x),\] and taking degrees, \[\eta_q(C)\leq \deg c(x)+\deg g(x)\leq k-r+ n-k=n-r.\]
\end{IEEEproof}

\subsection{Refinement of the Singleton bound for the Rank Weight of Cyclic Codes}
\begin{prop}
\label{p6} Let $C$ be an $[n,k]$ cyclic code over $\FF_{q^m}$ with $n$ and $q$ coprime. Then $d(\lambda(C))\leq \min(\eta_q(C^\perp)-k+1,m).$
\end{prop}

\begin{IEEEproof}
From Proposition \ref{p5}, $d_r(\lambda(C^{\perp}))=r$ if and only if $r\leq n-\eta_q(C^\perp)$. Hence, \[d_1(\lambda(C^\perp))=1,\ldots, d_{n-\eta_q(C^\perp)}(\lambda(C^\perp))=n-\eta_q(C^\perp).\] Now using Proposition \ref{p1} (\cite{D13}), 
\[\begin{aligned} \{d_r(\lambda(C))|& 1\leq r\leq k\} \subset \\&\{1,...,n\}  \setminus \{n+1-d_s(\lambda(C^\perp)) | 1\leq s\leq n-k\}.\end{aligned}\] Therefore, for every $1\leq r\leq k$, we have
\[d_1(\lambda(C)) < \cdots < d_k(\lambda(C))\leq \eta_q(C^\perp)+1.\]
Equivalently, $d(\lambda(C))=d_1(\lambda(C))\leq \eta_q(C^\perp)-k+1.$
\end{IEEEproof}


\begin{exmp}\rm
Let $C$ be the $[11,8]$-cyclic code over $\FF_{3^5}$ with generator polynomial $(x+1)(x+\alpha^2+\alpha-1)(x+\alpha^3+\alpha^2+\alpha)$, where $\alpha$ is a primitive element of $\FF_{3^5}$ over $\FF_3$ satisfying the equation $\alpha^5=\alpha+1$. Note here that $11$ divides $3^5-1$, so $x^{11}-1$ is split in $\FF_{3^5}[x]$. Then \[\begin{aligned} \eta_3(C) &=[\FF_3:\FF_3]+ [\FF_3(-\alpha^2-\alpha+1):\FF_3]\\ &\hspace{1cm}+[\FF_3(-\alpha^3-\alpha^2-\alpha):\FF_3]\\ &=1+5+5=11 \end{aligned}\] ($-\alpha^2-\alpha+1$ and $\alpha+1$ are not conjugate over $\FF_3$), so by Proposition \ref{p2}, we have $d(\lambda(C))>1$. Note that the Singleton bound gives that \[d(\lambda(C))\leq \min(n-k+1,m)=\min(11-8+1,5)=4.\] Taking now the dual code $C^\perp$, its generator polynomial is \[\begin{aligned}& g^\perp(x) =h(0)^{-1}x^8h(x^{-1})=\\ &(x+\alpha^2+\alpha-1)(x+\alpha^3+\alpha^2+\alpha)(x+\alpha^3+\alpha^2-\alpha-1)\\ &(x+\alpha^3+\alpha-1)(x+\alpha^4+\alpha^3+1)(x+\alpha^4-\alpha^3+1)\\ &(x-\alpha^4+\alpha^3+\alpha^2-\alpha-1)(x-\alpha^4-\alpha^3+\alpha^2+1),\end{aligned}\] with $h(x)=\frac{x^n-1}{g(x)}$. This yields that $\eta_3(C^\perp)=5+5=10$, so, by Proposition \ref{p6}, \[d(\lambda(C))\leq 10-8+1=3.\]
Finally, $d(\lambda(C))\in \{2,3\}.$ In fact, it is equal to $2$ here, the codeword \[[0, 0, 0, 0, 0, 0, 1, -\alpha^4 -\alpha^3 + 1, 0, 1, -\alpha^4 -\alpha^3 + 1]\] having rank weight $2$.  
\end{exmp}


%
%

\section*{Acknowledgement}
The research of J. Ducoat and F. Oggier is supported by the Singapore National Research Foundation under Research Grant NRF-RF2009-07.

%
%


\begin{thebibliography}{8}
%
\bibitem{Gabidulin}
Gabidulin, E. M. (1985) Theory of codes with maximal rank distance. Problems of Information Transmission, vol. 21, pp. 1--12.

\bibitem{OS12}
F. Oggier, A. Sboui, On the Existence of Generalized Rank Weights, {\em International Symposium on Information Theory and Its Applications  (ISITA 2012)}, Honolulu, Hawaii, 2012. 
%
\bibitem{KMU13}
J. Kurihara, R. Matsumoto, T. Uyematsu, ``Relative Generalized Rank Weight of Linear Codes and Its Applications to Network Coding", available at \url{http://arxiv.org/abs/1301.5482}.
%
\bibitem{D13}
J. Ducoat, ``Generalized rank weights: a duality statement", {\em Finite Fields and Applications (Fq11)}, 2013, available at \url{http://arxiv.org/abs/1306.3899}. 
%
\bibitem{SR03}
U. Sripati, B. Sundar Rajan, ``On the Rank-Distance of Cyclic Codes,''
{\em IEEE International Symposium on Information Theory (ISIT 2003)}, Yokohama, Japan,
2003. 
%
\bibitem{SR03-2}
U. Sripati, B. Sundar Rajan, ``On the Rank-Distance of Cyclic Codes,''
{\em Technical Report TR-PME-2003-04}, May 2003.
2003. 

\bibitem{S90}
H. Stichtenoth, ``On the Dimension of Subfield Subcodes,'' {\em IEEE Transactions on Information Theory}, vol. 36(1), 1990.
%
\end{thebibliography}
\end{document}